\documentclass[letterpaper, 10 pt, journal, twoside]{IEEEtran}
\usepackage{amsmath,amsfonts}
\usepackage{algorithmic}
\usepackage{algorithm}
\usepackage{array}
\usepackage{subfigure}
\usepackage{textcomp}
\usepackage{stfloats}
\usepackage{url}
\usepackage{verbatim}
\usepackage{graphicx}
\usepackage{cite}
\usepackage{color}
\usepackage{amsthm}
\usepackage{wrapfig}
\usepackage{caption}
\usepackage{tabularray}
\UseTblrLibrary{booktabs}
\usepackage[colorlinks,linkcolor=blue,citecolor=red,urlcolor=black,bookmarks=false,hypertexnames=true]{hyperref} 

\newtheorem{definition}{Definition}
\newtheorem{lemma}{Lemma}
\newtheorem{proposition}{Proposition}

\theoremstyle{definition}

\newcommand\norm[1]{\left\lVert#1\right\rVert}

\begin{document}

\title{Robust Lateral Control of a Convoy of Autonomous \& Connected Vehicles with Limited Preview}

\author{Mengke Liu$^{1}$, Neelkamal Somisetty$^{2}$, Sivakumar Rathinam$^{2}$, Swaroop Darbha$^{2}$

\thanks{$^{1}$ Mengke Liu is with State Key Laboratory of Advanced Automotive Integration and Control, FAW Group, Changchun, Jilin 130000, PRC (e-mail: {\tt\footnotesize liumengke@faw.com.cn}).$^{2}$ Swaroop Darbha, Neelkamal Somisetty, and Sivakumar Rathinam are with the Department of Mechanical Engineering, Texas A\&M University, College Station, TX 77843, USA (e-mail: {\tt\footnotesize  \{dswaroop, neelkamal.sept18, srathinam\}@tamu.edu}).}%
}

\maketitle
\begin{abstract}
This paper addresses the lateral control of Autonomous \& Connected Vehicles (ACVs) convoys during Emergency Lane Change (ELC) maneuvers. These maneuvers are initiated in response to emergency cues from either the front or rear of the convoy, responding to the need to avoid obstacles or facilitate the passage of other vehicles. The primary objective of this study is to develop a lateral control scheme for ACVs based on the available information. The foundational assumption in this study is the existence of reliable connectivity among ACVs, wherein each subsequent ACV possesses information concerning the GPS position traces of both the lead and immediately preceding vehicles within the convoy. This connectivity facilitates the construction of a composite ELC trajectory that synthesizes this information, serving as a “discretized” preview of the trajectory to be tracked. The procedural steps include constructing this composite trajectory, determining cross-track error, heading, and yaw rate errors relative to it, and subsequently formulating a lateral control strategy. Furthermore, the paper presents findings on the lateral string stability of ACV convoys across various scenarios, encompassing changes in longitudinal velocity and scenarios where lead vehicle information is unavailable. Numerical and experimental results validate the efficacy of the proposed lateral control scheme for ACV convoys.
\end{abstract}


\section{Introduction}
The lateral control of Autonomous and Connected Vehicles (ACVs) often relies on road infrastructure elements such as reference wires, magnets, embedded signals, or lane markers, as highlighted in prior research \cite{alleleijn2014lateral, peng1993magnetic, taylor1999comparative}. This reliance introduces limitations related to cost and susceptibility to adverse weather conditions, such as snow or smog. Additionally, in scenarios requiring close following distances, such as eco-driving, or on roads without lane markings, detecting lane markings may not be feasible. One such study \cite{theodosis2023cruise} focuses on the design of cruise controllers for lane-free driving environments, highlighting the challenges and solutions associated with operating without traditional road infrastructure. In defense applications, the availability of road infrastructure for convoying purposes may be limited \cite{liu2020lateral_ai}. During an Emergency Lane Change (ELC) maneuver, the utility of road infrastructure, such as lane markings or embedded magnets, may be reduced compared to its effectiveness in lane-keeping scenarios. Consequently, there is a need to develop lateral control schemes that operate independently of road infrastructure, leveraging the advantages of connectivity instead. 

The primary contribution of this paper is the development of lateral control algorithms for a convoy of ACVs that do not rely on road infrastructure, but instead utilize connectivity to execute ELC maneuvers. An ELC maneuver is typically triggered by emergency cues from the front or rear of the convoy in response to avoiding an obstacle or making way for other vehicles to pass. Connectivity is crucial for executing an ELC maneuver, as each ACV in the convoy possesses information about its position, heading, yaw rate in a ground frame, and longitudinal speed, which can be communicated to other ACVs. This connectivity enables each vehicle to access the position information of both the lead ACV and its immediate preceding ACV.
\begin{figure}[ht]
\begin{center}
  \includegraphics[scale=0.25]{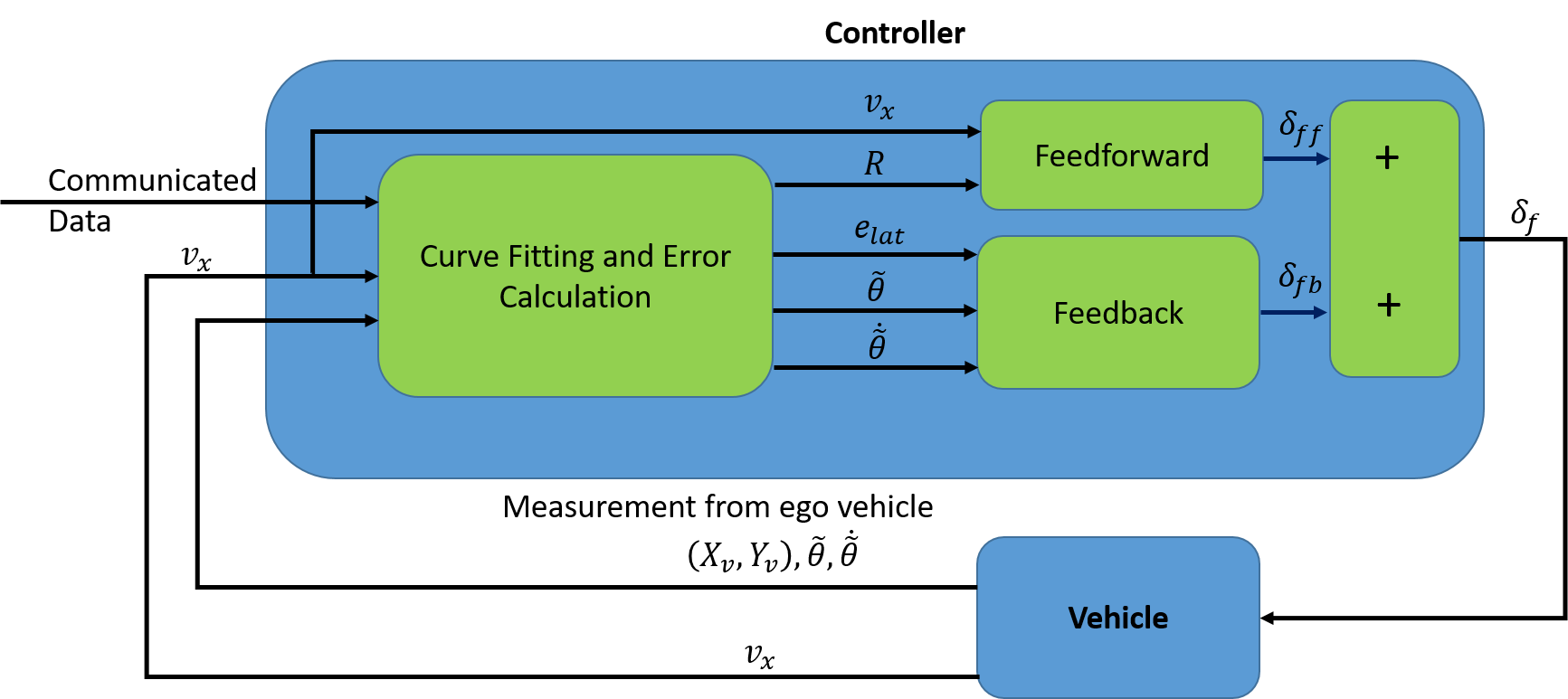}  
\end{center}
\caption{Structure of the proposed lateral controller.}
\label{fig:ControllerStructure}
\end{figure}
The main challenge for each following ACV encompasses three key aspects: (1) determining a ``target'' trajectory based on the available sensed and communicated information, (2) calculating the feedforward steering command by determining the curvature of the trajectory, and (3) computing error signals relative to the target trajectory, and generating a feedback control action to regulate the ACV's trajectory towards the target trajectory. Error signals used to assess controller performance include cross-track error (the distance between the ACV and the closest point on the target trajectory), heading error (the deviation between the ACV's heading and the direction of the tangent to the target trajectory at the closest point), and yaw rate error. A schematic representation of the proposed lateral controller architecture addressing these challenges is depicted in Fig. \ref{fig:ControllerStructure}.

\subsection{Relation to Literature \& Novelty of the Proposed Work}
A concise overview of lane-changing maneuvers and associated control approaches can be found in prior works \cite{yanumula2023optimal, badnava2021platoon, bevly2016lane, shladover1995review}. In contrast to the use of vision systems, magnetometers, or guided wires \cite{taylor1999comparative, lu2004practical, peng1993magnetic}, the proposed work solely relies on GPS/IMU measurements of the ego vehicle's position, heading, yaw rate, and longitudinal velocity  (as indicated in the outputs of the vehicle block in Fig. \ref{fig:ControllerStructure}), along with communicated position information from the lead and preceding vehicles. The concept of utilizing the information of preceding vehicle positions was explored in a previous study by Gehrig and Stein \cite{gehrig1998trajectory}. However, their work only considered a two-vehicle platoon with Ackerman steering input and acknowledged the need for refining the dynamic model. In contrast, the proposed work incorporates a dynamic model that accounts for lateral acceleration at high speeds and addresses the control of multiple-vehicle convoys. In \cite{zhao2021combined}, the study focused on a mixed-vehicle platoon consisting of ACVs and human-driven vehicles (HVs) on curved roads. Reference trajectories for following ACVs are constructed based on key points from the trajectories of detected HVs. Conversely, in this study, a composite ELC trajectory is constructed for the ACV from the communicated trajectory information of both the lead and preceding vehicles.

Based on the constructed target trajectory, error signals are computed as depicted in Fig. \ref{fig:ControllerStructure}. Previous studies \cite{lu2004practical, hassanain2020string, liu2020lateral_ai} have highlighted the potential for cross-track errors to amplify when each ACV relies solely on information from the preceding ACV, analogous to the longitudinal control case \cite{swaroop1996string}. In this paper, we provide proof that cross-track errors can amplify even when the lead vehicle is traveling in a straight line. To mitigate string instability, inter-vehicular communication has been proposed as a practical solution \cite{lu2004practical}. In this study, the lead ACV's position, heading, and yaw rate information are communicated to each following ACV in the convoy, similar to the longitudinal control architecture employed by California PATH \cite{swaroop1996string}.
Given that ACVs may have different characteristics, it may not be suitable to communicate vehicle-specific information such as throttle angle, steering angle, or brake pedal position. Therefore, from an implementation perspective, it is simpler to focus on communicating information related to vehicle kinematics. While the proposed architecture shares similarities with the longitudinal vehicle following architecture utilized by California PATH \cite{swaroop1996string, lu2004practical, hassanain2020string}, the specific details of how the lead vehicle information is utilized differ significantly.

Another relevant study by Lu et al. \cite{lu2004practical} emphasizes the importance of inter-vehicular communication in mitigating string instability. Their control structure differs due to variations in sensing methods, specifically aiming to match the position of a rigid extension of the ego vehicle's front bumper in the longitudinal direction to the position of the immediately preceding vehicle's back bumper. Another study on the lateral control of convoys \cite{alleleijn2014lateral} proposes the communication of lateral acceleration to alleviate string instability concerns by enforcing a condition on the propagation transfer function of the rate of change of the heading angle of the preceding vehicle. 

Existing literature on vehicle lateral control \cite{lu2004practical, tan1999development, fujioka1998vehicle, hassanain2020string} has predominantly modeled the closed-loop system as a linear time-invariant system. Some studies have accounted for the time-varying nature of longitudinal speed as an uncertainty, designing loop-shaping or \(\mathcal{H}_{\infty}\) controllers to enhance robustness in their control designs. In contrast, this paper adopts a frozen-parameter approach. It seeks to identify a stabilizing set of controllers for a constant longitudinal speed and determine a controller that lies within the intersection of stabilizing sets across a range of operational longitudinal speeds. Given that the longitudinal speed of an ACV can vary with time, albeit slowly, the stability of a closed-loop system with time-varying longitudinal speed becomes a pertinent question, which this paper addresses.


The contributions of this paper are significant and manifold. Firstly, we demonstrate lateral string instability with predecessor-only information, even when the lead vehicle is tracking a straight line. Motivated by this finding and the longitudinal control approach used in \cite{swaroop1996string}, we introduce a direct method for integrating information from both leading and preceding vehicles, facilitating the generation of target trajectories designed to enhance lateral string stability by reducing the propagation of cross-track errors. Secondly, we provide numerical corroboration of the lateral controller's effectiveness for a convoy of ACVs, building upon the feedback controller framework established in prior research \cite{liu2019lateral, liu2020lateral_ai}. Moreover, this study makes an additional contribution by demonstrating that employing a frozen-parameter controller design approach guarantees closed-loop stability under specific conditions: (a) the ACV's longitudinal speed surpasses a nonzero minimum threshold and does not exceed a defined upper limit, and (b) the longitudinal acceleration remains square integrable over time. These conditions are practically relevant given the finite duration of ELC maneuvers. In addition to theoretical analyses and numerical simulations, this paper extends its investigations to practical applications by implementing the proposed control algorithms on a Lincoln MKZ vehicle. This real-world application substantiates the string stability of the ACV when utilizing preview data from both lead and preceding ACVs, thereby underscoring the practical relevance and efficacy of the proposed lateral control strategy.

\subsection{Organization}
The structure of this paper is as follows: Section \ref{sec:model} offers a brief overview of the vehicle model. Section \ref{sec:design} introduces the proposed lateral control scheme. Section \ref{sec:robustness} discusses the robustness of the lateral string stability of the ACV convoys. Section \ref{sec:results} presents the simulation and experimental results. Finally, Section \ref{sec:conclusions} concludes the paper with final remarks.

\section{Mathematical Model for Lateral Dynamics of an ACV}
\label{sec:model}

The standard ``bicycle'' model has been the primary basis for lateral vehicle control \cite{kapania2015design, karafyllis2022constructing, liu2019lateral, hassanain2020string}. Let \(m_v\) and \(I_z\) represent the vehicle mass and moment of inertia about its center of mass, respectively. Let \(C_f\) and \(C_r\) denote the cornering stiffness of the front and rear axles, respectively, and \(a\) and \(b\) the distances from the center of gravity to the front and rear axles, respectively. For a front-steered vehicle with steering input \(\delta_f\), applying Newton-Euler's equations along with linear constitutive equations for cornering forces from the tires yields the following equations of motion:

\begin{eqnarray}
m_v(\frac{dv_y}{dt}+v_x\dot{\theta})=C_f\delta_f-\frac{C_f+C_r}{v_x}v_y-\frac{aC_f-bC_r}{v_x}\dot{\theta},\label{ForceEOM}\\
    I_z\ddot{\theta}=aC_f\delta_f - \frac{aC_f-bC_r}{v_x}v_y-\frac{a^2C_f+b^2C_r}{v_x}\dot{\theta}.\label{MomentEOM}
\end{eqnarray}

\begin{figure}
\centering
\includegraphics[scale=0.4]{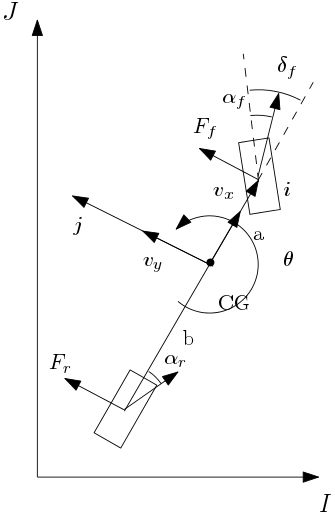}
\caption{Illustration of the bicycle model.}
\label{fig:bicycle}
\end{figure}

For the design of a feedback controller, it is desired to recast the equations of motion using errors in position and heading with respect to the desired trajectory. Based on previous work \cite{liu2019lateral, liu2020lateral}, we can express the vehicle dynamic equations (\ref{ForceEOM}) and (\ref{MomentEOM}) in terms of the states \(e_{lat}, \dot{e}_{lat}, \tilde{\theta}, \frac{d\tilde{\theta}}{dt}\) using the following matrices:
\begin{eqnarray}
\mathbf{M} &:= &\begin{bmatrix}
m_v & 0 \\
0 & I_z
\end{bmatrix}, \quad
\mathbf{C} \ := \begin{bmatrix}
\frac{C_f+C_r}{V_0} & \frac{aC_f-bC_r}{V_0} \\
\frac{aC_f-bC_r}{V_0} & \frac{a^2C_f+b^2C_r}{V_0}
\end{bmatrix},\nonumber \\
\mathbf{B} &:= &\begin{bmatrix}
1 \\
a 
\end{bmatrix}, \quad
\mathbf{F} \ := \begin{bmatrix}
m_vV_0^2+(aC_f-bC_r) \\
a^2C_f+b^2C_r
\end{bmatrix},\nonumber \\
\mathbf{L} &:= &\begin{bmatrix}
0 & -(C_f + C_r) \\
0 & -(aC_f - bC_r)
\end{bmatrix}, \quad 
\mathbf{x} \ := \begin{bmatrix}
e_{lat} \\
\tilde \theta
\end{bmatrix}.\nonumber 
\end{eqnarray}
The physical meaning of the states, as well as their computation (except for \(\dot{e}_{lat}\)), based on available measurements, is provided in the next section. 

The governing equations can thus be written as:
\begin{eqnarray}
\mathbf{M} \ddot {\mathbf{x}} + \mathbf{C} \dot {\mathbf{x}} + \mathbf{L} \mathbf{x}= \mathbf{B} C_f \delta_f - \mathbf{F} (\frac{1}{R}).
\label{eq:System}
\end{eqnarray}
Based on our previous work \cite{liu2020lateral}, we use an experimentally corroborated second-order model for steering actuation dynamics; a similar model was used in \cite{hassanain2020string}. The actuation transfer function is given by:

\begin{align}
       H_a(s) = \frac{w_n^2}{s^2+2 \zeta w_ns+w_n^2},
       \label{Eq:Transfer_fn}
\end{align}
where
\begin{align*}
    w_n = \sqrt{\frac{K_r}{J_w}}, \ \zeta = \frac{b_w}{\sqrt{K_rJ_w}}.
\end{align*}
In the above equation, \(J_w\) represents the steering wheel inertia, \(b_w\) is the torsional viscous damping coefficient, and \(K_r\) is the torsional stiffness of the steering column. The input to the transfer function is the steering angle command. Below is the table of parameters we have determined experimentally \cite{liu2020lateral, liu2019lateral} for the Lincoln MKZ vehicle, a passenger mid-size sedan. These parameters are used in our simulation study. In this table, \(m_v = \frac{W}{g}\).

\begin{table}[h!]
  \begin{center}
\caption{Vehicle parameters}
    \label{tab:table1}
    \resizebox{0.5\textwidth}{!}{\begin{tblr}{l|l|r|l} 
    \hline
      \textbf{Parameter} & \textbf{Description} & \textbf{Value }& \textbf{Unit} \\ 
      \hline 
      $m_v$ & Vehicle total mass & 1896 &[$kg$]\\
      $W_f$ & Vehicle front axle load  & 1052.32g &[$N$]\\
      $W_r$ & Vehicle rear axial mass & 843.68g &[$N$]\\
      $I_z$ & Vehicle inertia & 3803 & [$kg.m^2$]\\
      $C_f$ & Cornering stiffness of front axle & 400000 &[$N/rad$]\\
      $C_r$ & Cornering stiffness of rear axle & 381900 &[$N/rad$]\\
      $a$ & Distance of c. g. to the front axle & 1.2682 & [$m$]\\
      $b$ & Distance of c. g. to the rear axle& 1.5818 & [$m$]\\
       $\zeta $& Damping ratio of steering actuation &0.4056 & [-]\\
      $\omega_n $& Natural frequency of steering actuation & 21.4813&[$rad/s$]\\
      \hline
    \end{tblr}}
  \end{center}
\end{table}

Through experimental analysis, we determined the values \(\zeta = 0.4056\) and \(\omega_n = 21.4813 \ \text{rad/s}\). These values can be compared with the values \(\zeta = 0.7\) and \(\omega_n \approx 19 \ \text{rad/s}\) employed in the study by Hassanain et al. \cite{hassanain2020string}. We use these values for the design and implementation of the lateral controller.

\section{Lateral Controller Design}
\label{sec:design}

This section outlines the design of the lateral controller, depicted in Fig. \ref{fig:ControllerStructure}. The design process involves the following key tasks: 1. Construction of the target trajectory. 2. Calculation of curvature and feedback error signals relative to the target trajectory. 3. Computation of the feedforward controller. 4. Computation of the feedback controller.

\subsection{Target Trajectory Construction}
\label{section:Target Trajectory Construction}

\begin{figure}[ht]
    \centering
    \includegraphics[scale=0.4]{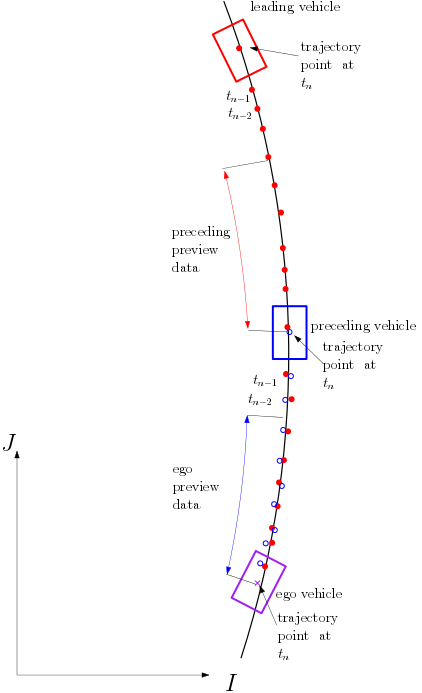}
    \caption{Illustration of the sampled trajectories of ACVs in the convoy and the information available to the ego ACV.}
    \label{fig:leadandfollower}
\end{figure}

To reiterate \cite{lu2004practical, liu2020lateral_ai}, the consideration of GPS information from lead and preceding vehicles arises from the need to address string instability in the lateral direction, ensuring that cross-track errors do not amplify to maintain safety. Figure \ref{fig:leadandfollower} illustrates a convoy where the ego ACV (shown in purple) has access to sampled position data from the ACVs ahead. The red and blue dots correspond to the locations where their position data was transmitted to other ACVs, including the ego ACV. The ego ACV receives and stores this data, using only the data within a distance \(L_{preview}\) ahead of it. This is shown in Figure \ref{fig:leadandfollower} as the ego preview data, which can be sorted in increasing order of distance from the ego ACV.

Let \((x_1^l, y_1^l), (x_2^l, y_2^l), \ldots, (x_N^l, y_N^l)\) (corresponding to red dots in the ego preview data) denote the GPS data of the lead vehicle. Similarly, let \((x_1^p, y_1^p), (x_2^p, y_2^p), \ldots, (x_M^p, y_M^p)\) (corresponding to blue dots in the ego preview data) denote the GPS data of the preceding ACV. The problem is to construct a target trajectory as a circular arc spline approximation of the data at hand. A circular arc spline\footnote{A circular arc spline is a union of straight line segments and circular arcs, resulting in a continuous curve with a defined tangent everywhere but allowing for piecewise constant curvature.} is particularly useful because (1) a majority of U.S. roads are built as circular arc splines \cite{cheu2006highway}, and (2) computation of curvature and error signals (namely, cross-track, heading, and yaw rate errors) becomes simpler. Identifying the points of inflection of the circular spline is crucial to ensuring that the ego vehicle takes appropriate steering action, and this issue is addressed in the next paragraph. 

As mentioned in \cite{alleleijn2014lateral, taylor1999comparative}, human drivers require preview or lookahead information equivalent to about \(0.8 \; \text{seconds}\) of time headway. In other words, if \(v\) is the ACV's longitudinal speed, then preview data within \(L_{preview} = 0.8 \times v \approx 30 \; \text{meters}\) should suffice for highway speeds. Notably, the maximum frequency of GPS data updates is typically constrained to \(20 \; \text{Hz}\). Consequently, the number of preview samples for both preceding and lead vehicles, denoted as \(M\) and \(N\) respectively, is subject to the limitation \(M, N < 20\). 

The duration for a lane change is typically \(5-6 \; \text{seconds}\) \cite{toledo2007modeling}. Therefore, it is reasonable to assume that there can be at most one change in the curvature of the target trajectory within \(1 \; \text{second}\). We currently use GPS-RTK, which has an accuracy of approximately \(2 \; \text{cm}\) in position; thus, the transmitted data is assumed to have the same order of accuracy. Given that the radius of curvature of the data is typically on the order of hundreds of meters, and the distance between successive samples of GPS data transmitted by the ACVs is on the order of \(1 \; \text{meter}\), it suffices to consider circular arc splines where circular arcs and straight lines alternate.

Since the data is sorted and the GPS-RTK provides position data with centimeter accuracy, with the distance between successive data samples from the ACV being on the order of meters, we can exploit this situation further. For example, consider the case where the ego ACV, traveling straight, needs to decide whether to turn based on the available preview data. Since the data is sorted by distance, consider the data from the preceding vehicle (the same can be done for the lead vehicle as well): Find the largest \(k \geq 3\) such that points \((x_2^p, y_2^p), \ldots, (x_{k-1}^p, y_{k-1}^p)\) are no farther than a threshold, say \(\epsilon \; \text{cm}\), from the line connecting \((x_1^p, y_1^p)\) and \((x_k^p, y_k^p)\). Clearly, we can associate a straight line segment with the data \((x_1^p, y_1^p), \ldots, (x_k^p, y_k^p)\) and associate a circular arc with the rest of the data. We choose the threshold to be \(10 \; \text{cm}\) (about five times the accuracy specified by GPS-RTK). This allows us to focus on the circular arc approximation of the remaining data.

Leveraging the preview data obtained from both the lead ACV and the preceding ACV, we systematically synthesize a composite ELC trajectory. This trajectory is designed to amalgamate the GPS information from both the lead and preceding vehicles for every subsequent vehicle within the convoy. In the following subsection, we will provide a comprehensive formulation for fitting a circular arc.

\subsubsection{Fitting a circular arc based on preview data}
There are two sources for the preview data—the lead ACV and the preceding ACV—one may want to weigh the data differently. For this purpose, consider a weighing parameter \(\alpha \in [0,1]\) that will be used in defining a convex combination of two different error functions.

Let \((X_c, Y_c)\) be the coordinates of the center of the desired circle and \(R\) be its radius. A circle is defined by an equation of the form: 
\[ e(x,y) = 0, \]
where 
\[ e(x,y) := (x-X_c)^2 + (y-Y_c)^2 - R^2. \]
One can interpret \(\pi e\) as the difference between the areas of a circle passing through a point \((x,y)\) with a center at \((X_c, Y_c)\) and another circle of radius \(R\). Associated with the given data \((x_1^p, y_1^p), \ldots, (x_M^p, y_M^p)\) and \((x_1^l, y_1^l), \ldots, (x_N^l, y_N^l)\), one can define a composite error as

\begin{equation}
    J = \alpha \sum_{i=1}^M (e(x_i^p, y_i^p))^2 + (1-\alpha) \sum_{j=1}^N (e(x_j^l, y_j^l))^2.
\end{equation}

Then, \((X_c, Y_c)\) and \(R\) can be determined as the arguments that minimize \(J\) and can be derived from three linear equations in \(X_c, Y_c,\) and \((R^2 - X_c^2 - Y_c^2)\).

In this study, we select \(\alpha = 0.5\), signifying an equal allocation of weight to the data from both the lead and preceding ACVs. This weighting is used to determine the composite ELC trajectory. The choice of \(\alpha = 0.5\) simplifies implementation, as it allows us to treat the data from both the lead and preceding ACVs as though it originated from a single source.

Once an ELC trajectory is computed, it is necessary to calculate the feedback signals: the lateral error of the vehicle from the trajectory (\(e_{lat}\)), the heading error of the vehicle (\(\tilde{\theta}\)), and the yaw rate error (\(\dot{\tilde{\theta}}\)). The radius of curvature allows us to compute the feedforward control with respect to the ELC trajectory, and the feedback signals help the vehicle track the trajectory.

\subsection{Computation of Feedback Errors}
Let \((X_v, Y_v)\) denote the ego ACV's position. The computation of feedback error signals depends on whether the portion of the trajectory represented by preview data is a straight line segment or a circular arc.

Given the provided preview data, the nature of the trajectory can be discerned. A simple heuristic is employed: utilizing the timestamped and sequenced nature of the data, we connect the initial sample \((x_1, y_1)\) and the final sample \((x_N, y_N)\) with a straight line segment, denoted as \(\mathcal{L}\). For each \(i\) ranging from 1 to \(N\), the distance (\(dis_i\)) of the samples \((x_i, y_i)\) from \(\mathcal{L}\) is computed using the formula:
\[ dis_i = \left|\frac{(y_i - y_1)(x_N - x_1) - (x_i - x_1)(y_N - y_1)}{\sqrt{(x_N - x_1)^2 + (y_N - y_1)^2}}\right|. \]
If \(\max_i dis_i < 0.1 \; \text{m}\), the preview data is considered to represent a straight line segment. Otherwise, a circular arc is fitted through the given data.

\begin{figure}
\centering
\includegraphics[scale=0.3]{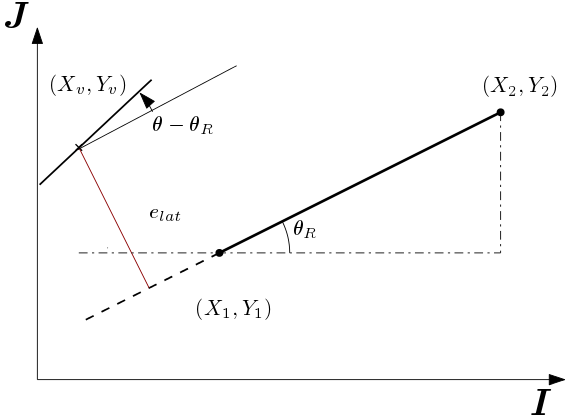}
\caption{Heading and lateral errors for a straight line segment.}
\label{fig:line}
\end{figure} 

In the case of a straight-line trajectory, the desired yaw rate is zero. As illustrated in Fig. \ref{fig:line}, the lateral error can be computed as follows:
\[ e_{lat} = \frac{Y_v - mX_v - c}{\sqrt{1 + m^2}}, \]
where \( y = mx + c \) is the equation of the straight line based on the closest point \((X_1, Y_1)\) and the second closest point \((X_2, Y_2)\). The desired yaw angle \(\theta_R\) is also determined based on these two points. The heading error is given by (see Fig. \ref{fig:line}):
\[ \tilde{\theta} = \theta - \theta_R, \]
and the yaw rate error is given by:
\[ \dot{\tilde{\theta}} = \dot{\theta}. \]

If part of the trajectory is a circular arc, the projection of \((X_v, Y_v)\) onto the circular arc can be constructed by drawing a line from the center of the circle \((X_c, Y_c)\) to the current position of the vehicle \((X_v, Y_v)\) and extending it, if necessary, to meet the circular arc at \((X_0, Y_0)\). The angle made by the tangent at \((X_0, Y_0)\) with the positive \(X\) axis of the ground is represented by \(\theta_R\). As shown, unit vectors \(i\) and \(j\) are attached to the vehicle along the longitudinal and lateral axes, respectively. The angle \(\theta\) made by unit vector \(i\) with unit vector \(I\) represents the vehicle's heading in the governing equations of motion.

\begin{figure}
\centering
\includegraphics[scale=0.3]{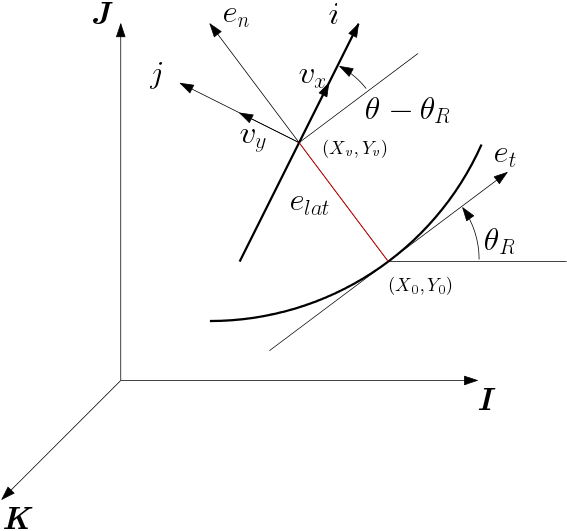}
\caption{Heading and lateral errors for a circular arc.}
\label{fig:circle}
\end{figure}

The lateral error can be computed as follows (see Fig. \ref{fig:circle}):
\[ e_{lat} = R - \sqrt{(X_v - X_c)^2 + (Y_v - Y_c)^2}. \]
The heading error can similarly be computed as (see Fig. \ref{fig:circle}):
\[ \tilde{\theta} = \theta - \theta_R, \]
where \(\theta_R\) is defined based on the closest point and the second closest point in the preview section. Correspondingly, the yaw rate error is given by:
\[ \dot{\tilde{\theta}} = \dot{\theta} - \frac{v_x}{R}. \]

\subsection{Lateral Controller Synthesis}
The lateral controller for tracking a trajectory can be decomposed into two parts: a feedforward component and a feedback component, i.e.,
\[ \delta_c = \delta_{ff} + \delta_{fb}. \]
In the following subsections, we will outline the design of these controller components.

\subsection{Feedforward Controller Design}
The feedforward controller essentially provides the steering input that would keep an ACV on a circular trajectory without any feedback if the initial conditions were appropriate and if no disturbance inputs were present. One may view the feedforward steering input, \(\delta_{ff}\), as the steering input that results in the ACV's trajectory being a circle of radius \(R\) when it is traveling at a longitudinal speed of \(V_0\). The corresponding feedforward steering input is defined as follows:
\begin{equation}
    \delta_{ff} = \frac{a+b}{R} + K_{sg} \frac{V_0^2}{R},
\end{equation}

where $K_{sg}$ is the steer gradient. It is important to note that for vehicles exhibiting understeer, where \( K_{sg} > 0 \), the open-loop handling dynamics are inherently stable \cite{segel1964investigation, crolla2012impact}. This relationship can be derived by ensuring that \(\ddot{x}\) and \(\dot{x}\) are zero. We leverage this relationship to derive key parameters of the vehicle's dynamic model. For a comprehensive discussion on this topic, please refer to \cite{liu2020lateral}.

\subsection{Feedback Controller Design}
Feedback control is based on the available information, namely, the errors \(e_{lat}\), \(\tilde{\theta}\), and \(\dot{\tilde{\theta}}\) with respect to the composite ELC trajectory. Since lateral velocity information is not readily available, we seek a control law of the form:

\begin{equation}
    \delta_{fb} = -k_e e_{lat} - k_{\theta} \tilde{\theta} - k_w \dot{\tilde{\theta}},
\end{equation}

where the gains \(k_e\), \(k_{\theta}\), and \(k_w\) need to be determined. 

\subsection{Construction of the Set of Stabilizing Structured Feedback Controllers}
The selection of gains \(k_e\), \(k_{\theta}\), and \(k_{\omega}\) is crucial for the implementation of the lateral control system. We utilize the D-decomposition method, previously employed in our work \cite{liu2019lateral, liu2020lateral_ai}, to synthesize the set of stabilizing feedback gains. From the governing equations, the transfer functions relating \(e_{lat}(s)\) and \(\tilde{\theta}(s)\) to \(\delta_f(s)\) can be related through the open-loop characteristic polynomial, \(\Delta_o(s)\), as follows:

\begin{eqnarray*}
\Delta_o(s) &=& s^2 \left( m_v I_z s^2 + \frac{(I_z + m_va^2)C_f + (I_z + m_vb^2)C_r}{V_0} s \right. \\
&& \left. + \frac{(a + b)^2 C_f C_r}{V_0^2} - m_v(a C_f - b C_r) \right), \\
\frac{e_{lat}(s)}{\delta_f(s)} &=& \frac{C_f}{\Delta_o(s)} \left( I_z s^2 + \frac{b(a + b)C_r}{V_0} s + (a + b) C_r \right), \\
\frac{\tilde{\theta}(s)}{\delta_f(s)} &=& \frac{C_f}{\Delta_o(s)} \left( m_v a s^2 + \frac{(a + b)C_r}{V_0} s \right).
\end{eqnarray*}

Since, 
\begin{eqnarray}
\nonumber
 \delta_f &=& H_a(s) \delta_c(s) \\
 &=& - H_a(s)(k_e e_{lat}(s) + (k_{\theta}+k_ws) \tilde \theta(s))\nonumber,
\end{eqnarray}
it follows that the closed-loop characteristic equation can be expressed as:
\begin{eqnarray}
\Delta_o(s) + H_a(s) k_e C_f \left( I_z s^2 + \frac{b(a + b)C_r}{V_0} s + (a + b) C_r \right) \nonumber \\
+ H_a(s) (k_{\theta} + k_w s) C_f \left( m_v a s^2 + \frac{(a + b)C_r}{V_0} s \right) = 0.
\end{eqnarray}
For the second-order steering actuation model, it follows that the characteristic polynomial may be expressed in terms of the control parameter vector  $\mathbf{K} = (k_e, k_{\theta}, k_w)$ as:
\begin{eqnarray}
\nonumber
\Delta(s; \mathbf{K}) \hspace{-0.25em}&=& \hspace{-0.25em}(s^2+2\zeta\omega_n s+\omega_n^2) \Delta_o(s) \\
\nonumber
&&\hspace{-0.25em}+ k_e C_f \omega_n^2 \left( I_z s^2 + \frac{b(a + b)C_r}{V_0} s + (a + b) C_r \right) \\
\nonumber
&&\hspace{-0.25em}+ (k_{\theta} + k_w s) C_f \omega_n^2 \left( m_v a s^2 + \frac{(a + b)C_r}{V_0} s \right).
\label{eq:cl_poly}
\end{eqnarray}

Hence, a natural problem regarding the stability of tracking is to determine the set of control gains, \(\mathbf{K}\), such that the closed-loop characteristic polynomial in equation (\ref{eq:cl_poly}) is Hurwitz. An advantage of determining the entire set is that it can be pruned to accommodate additional performance criteria.

\textbf{Construction of the set of stabilizing fixed structure controllers:}
There are various methods to construct the set of controllers \cite{malik2008linear, bhattacharyya2018linear, henrion2003positive, siljak1988parameter} in the parameter space. Here, we adopt the D-decomposition approach from \cite{bhattacharyya2018linear, siljak1988parameter}. This approach relies on the continuous dependence of the roots of a polynomial on its coefficients for every regular perturbation. The method involves decomposing the parameter space into disjoint signature-invariant regions by identifying their boundaries. These boundaries can be obtained by (a) determining the set of \(\mathbf{K}\) for which \(\Delta(0, \mathbf{K}) = 0\) and (b) determining the set of \(\mathbf{K}\) for which \(\Delta(jw, \mathbf{K}) = 0\) for some \(w\). Once the parameter space is partitioned, each partition can be sampled to determine which corresponds to \(\Delta(s; \mathbf{K})\) being Hurwitz. 

Note that \(\Delta(s, \mathbf{K}) = A_6 s^6 + A_5 s^5 + A_4 s^4 + A_3 s^3 + A_2 s^2 + A_1 s + A_0\), where 
\begin{eqnarray*}\label{CharacteristicEquation}
A_6 &=& \frac{I_zm_v}{\omega^2_n}, \\
A_5 &=& \frac{2I_zm_v\zeta}{\omega_n} + \frac{C_f(I_z + a^2m_v) + C_r(I_z + b^2m_v)}{V_0 \omega^2_n}, \\
A_4 &=& \frac{2\zeta(C_f(I_z + a^2m_v) + C_r(I_z + b^2m_v))}{V_0 \omega_n} \\
     && + \left(\frac{(a + b)^2 C_f C_r}{V^2_0 \omega^2_n} - \frac{m_v(aC_f - bC_r)}{\omega_n^2}\right) + m_vI_z, \\
A_3 &=& \frac{2\zeta}{\omega_n}\left(\frac{(a + b)^2 C_f C_r}{V^2_0} - m_v(aC_f - bC_r)\right) \\
     && + \frac{C_f(I_z + a^2m_v) + C_r(I_z + b^2m_v)}{V_0} + C_f m_va k_{\omega}, \\
A_2 &=& \left(\frac{(a + b)^2 C_f C_r}{V^2_0} - m_v(aC_f - bC_r)\right) \\
     && + \left(C_f I_z k_e + C_f C_r \frac{(a + b)}{V_0} k_{\omega} + m_va C_f k_{\theta}\right), \\
A_1 &=& C_f C_r \frac{(a + b)}{V_0}(b k_e + k_{\theta}), \\
A_0 &=& C_f C_r (a + b) k_e,
\end{eqnarray*}
where \(\zeta\) and \(\omega_n\) are the damping ratio and the natural frequency of steering actuation, respectively, as found in our previous work \cite{liu2019lateral, liu2020lateral}.

It is important to note that the set of stabilizing controllers is specific to the operating longitudinal speed \(V_0\), as the coefficients \(A_0, A_1, A_2, A_3, A_4, A_5, A_6\) depend quadratically on \(\frac{1}{V_0}\). Define a parameter \(\gamma := \frac{1}{V_0}\); one can then construct the set of stabilizing controllers for each \(\gamma \in \left[\frac{1}{V_{max}}, \frac{1}{V_{min}}\right]\), where \(V_{min}\) and \(V_{max}\) are the limiting values of the operating longitudinal speeds. We then choose a control gain vector \(\mathbf{K} = (k_e, k_{\theta}, k_{\omega})\) that lies in the intersection of stabilizing sets as \(\gamma\) varies from \(\frac{1}{V_{max}}\) to \(\frac{1}{V_{min}}\).

\section{Robustness in Stability}
\label{sec:robustness}
In the construction of the set of stabilizing controllers, \(\gamma\) has been considered as a parameter that remains constant. However, the governing equations were simplified based on the assumption of a constant longitudinal speed. If the rate of change of longitudinal speed is sufficiently small, it is intuitive that guarantees of closed-loop stability will continue to hold based on the techniques of slowly varying parameter systems, as described in Section 9.3 of \cite{khalil2002nonlinear}. We will provide a similar bound here on the longitudinal acceleration of a vehicle performing the ELC to ensure that the closed-loop stability guarantee holds.

While the choice of a stabilizing feedback controller from the constructed set in the previous subsection ensures lateral stability with a constant longitudinal velocity and fixed vehicle mass, there is a need for robustness in stability to variations in these parameters.

\subsection{Closed-Loop Stability with Varying Longitudinal Velocity}

Let \(\mathbf{Q}\), \(\mathbf{G}\), and \(\mathbf{H}\) be appropriate matrices (in controllable canonical form) so that the transfer function \(H_a(s)\) from the steering command \(\delta_c\) to the steering angle \(\delta_f\) has the following minimal realization, with a corresponding state vector \(\eta\) of the steering actuation system:
\[ \dot{\eta} = \mathbf{Q} \eta + \mathbf{G} \delta_c, \quad \delta_f = \mathbf{H} \eta. \]

The dynamics of the vehicle can be expressed as:
\begin{eqnarray*}
\begin{pmatrix}
\dot{\mathbf{x}} \\ 
\ddot{\mathbf{x}} 
\end{pmatrix} 
= \begin{pmatrix} 
\mathbf{0} & \mathbf{I} \\ 
- \mathbf{L} & -\gamma \mathbf{C}_0 
\end{pmatrix} 
\begin{pmatrix}
\mathbf{x} \\ 
\dot{\mathbf{x}} 
\end{pmatrix} 
+ \begin{pmatrix}
0 \\ 
\mathbf{B} 
\end{pmatrix} C_f \delta_f 
- \begin{pmatrix}
0 \\ 
\mathbf{F} 
\end{pmatrix} \frac{1}{R}.
\end{eqnarray*}
Let \(\mathbf{K}_p := [k_e \; \; k_{\theta}]\) and \(\mathbf{K}_v := [0 \; \; k_{\omega}]\). Since 
\[ \delta_c = -\mathbf{K}_p \mathbf{x} - \mathbf{K}_v \dot{\mathbf{x}}, \]
the closed-loop dynamics is then given by:
\[
\underbrace{\begin{pmatrix}
\dot{\mathbf{x}} \\ 
\ddot{\mathbf{x}} \\ 
\dot{\eta}
\end{pmatrix}}_{\dot{\mathbf{z}}} 
= \underbrace{\begin{pmatrix}
\mathbf{0} & \mathbf{I} & \mathbf{0} \\
-\mathbf{L} & -\gamma \mathbf{C}_0 & \mathbf{BH} C_f \\
-\mathbf{G K}_p & -\mathbf{G K}_v & \mathbf{Q}
\end{pmatrix}}_{\mathbf{A}(\gamma)}
\underbrace{\begin{pmatrix}
\mathbf{x} \\ 
\dot{\mathbf{x}} \\ 
\eta
\end{pmatrix}}_{\mathbf{z}} 
+ \underbrace{\begin{pmatrix}
0 \\ 
-\mathbf{F} \\ 
0 
\end{pmatrix}}_{\tilde{\mathbf{F}}} \frac{1}{R}.
\]
 Once the controller gains are fixed, i.e., ${\bf K}_p, {\bf K}_v$ are known, closed-loop stability can be examined by  formally considering a linear system $\dot {\bf z} = {\bf A}(t) {\bf z}(t)$. Problem (9.31) of \cite{khalil2002nonlinear} can be employed to establish closed-loop stability; it is stated here for completeness:
 
\begin{lemma}
 Suppose  the linear system: $$\dot {\bf z} = {\bf A}(t) {\bf z}(t)$$
 satisfies the following conditions:
 \begin{itemize}
     \item For some $k>0$, $\|A(t)\| \le k$ for all $t \ge 0$, 
     \item For some $\sigma >0$, every eigenvalue of ${\bf A}(t)$ has a real part less than $-\sigma$ for every $t \ge 0$, and 
     \item $\|\dot {\bf A}(t)\|$ is square integrable, i.e., for some $\rho >0$, 
     $$\int_0^{\infty}\|\dot {\bf A}(t)\|^2 dt \le \rho. $$
  \end{itemize} 
  Then, the solution ${\bf z} = 0$ is exponentially stable. 
\end{lemma}
\begin{proof} Please see \cite{khalil2002nonlinear}. \end{proof}
  
Closed-loop stability with varying longitudinal speed has not been extensively considered in the literature. The following result establishes the connection between ``frozen-parameter'' control synthesis and the stability of the linear time-varying system at hand with varying longitudinal speed:
 
\begin{proposition}
Consider the linear time-varying system 
\[ \dot{\mathbf{z}} = \mathbf{A}(\gamma(t)) \mathbf{z} \]
given above. Suppose \(\mathbf{K}_p\) and \(\mathbf{K}_v\) have been chosen such that the real part of every eigenvalue of \(\mathbf{A}(\gamma)\) is less than \(-\sigma\) for every \(\gamma \in \left[\frac{1}{V_{max}}, \frac{1}{V_{min}}\right]\).

\begin{itemize}
    \item If the longitudinal acceleration of the vehicle is square integrable, i.e., 
    \[ \int_0^{\infty} \dot{v}_x^2(\tau) \, d\tau < \infty, \]
    then the equilibrium \(\mathbf{z} = 0\) is exponentially stable.

    \item If for some \(\mu, \alpha_0 \ge 0\), one of the following conditions holds:
    \begin{itemize}
        \item[(i)] 
        \[ \int_t^{t+T} \|\dot{v}_x(\zeta)\| \, d\zeta \le \mu T + \alpha_0, \quad \forall \, T, t \ge 0, \]
        \item[(ii)] 
        \[ \int_t^{t+T} \|\dot{v}_x(\zeta)\|^2 \, d\zeta \le \mu^2 T + \alpha_0, \quad \forall \, T, t \ge 0, \]
        \item[(iii)] 
        \[ \|\dot{v}_x(t)\| \le \mu, \]
    \end{itemize}
\end{itemize}
then there exists a \(\mu^*\) such that for all \(0 < \mu < \mu^*\), the equilibrium \(\mathbf{z} = 0\) is exponentially stable.
\label{prop:1}
\end{proposition}
\begin{proof}
    We will use the proof of Problem (9.31) of \cite{khalil2002nonlinear} to arrive at the result. Essentially, it requires that three conditions be satisfied: 
 \begin{itemize}
     \item \(\| \mathbf{A}(\gamma(t)) \|\) is bounded for all \(t\) or equivalently, it is bounded for all \(\gamma \in [ \frac{1}{V_{max}}, \frac{1}{V_{min}} ]\). This condition holds as \(\gamma(t) = \frac{1}{v_x(t)}\).
     \item  The real part of the eigenvalues of \(\mathbf{A}(\gamma(t))\) is less than \(-\sigma\) for some \(\sigma > 0\). This condition holds because the gains \(\mathbf{K}_p\) and \(\mathbf{K}_v\) have been chosen to be in the set of stabilizing controllers that render \(\mathbf{A}(\gamma(t))\) Hurwitz for every \(\gamma \in [ \frac{1}{V_{max}}, \frac{1}{V_{min}} ]\).
     \item The third condition requires \(\int_{0}^{\infty} \| \dot{\mathbf{A}}(\gamma(\tau)) \|^2 d \tau < \infty\). In our case, \(\mathbf{A}(\gamma)\) is linear in \(\gamma\), and for some appropriate constant matrices \(\mathbf{A}_0\) and \(\mathbf{A}_1\), we can express \(\mathbf{A}(\gamma) = \mathbf{A}_0 + \gamma \mathbf{A}_1\). Consequently,
\[ \dot{\mathbf{A}}(\gamma) = \dot{\gamma} \mathbf{A}_1, \]
implying that \(\dot{\mathbf{A}}(\gamma)\) is square integrable if \(\dot{\gamma}\) is square integrable, i.e., \(\dot{\gamma} \in \mathcal{L}_2\) as \(\gamma\) is bounded. 

However,
\begin{eqnarray*}
\dot{\gamma} &=& - \frac{1}{v_x^2} \dot{v}_x, \\
\implies \int_0^{\infty} | \dot{\gamma}(\tau) |^2 d \tau &=& \int_0^{\infty} \frac{1}{v_x^4(\tau)} | \dot{v}_x(\tau) |^2 d \tau \\
&\le& \frac{1}{V_{min}^4} \int_0^{\infty} | \dot{v}_x(\tau) |^2 d \tau.
\end{eqnarray*}
By virtue of \(\dot{v}_x\) being square-integrable, the third condition is also satisfied. Hence, the closed-loop time-varying system is exponentially stable by Lemma 1. Proof of the other parts of the theorem can be deduced from the proof of Theorem 3.4.11 (page 124) in Ioannou and Sun \cite{ioannou1996robust}.
\end{itemize}   
\end{proof} 
\vspace{-1em}
The assumption that the longitudinal acceleration is square integrable, bounded, or that the area is bounded linearly with the duration of the observation is very reasonable. This is because acceleration and deceleration maneuvers are typically associated with a finite change in longitudinal speed. The finite duration of these acceleration maneuvers allows this condition to be readily satisfied.

\subsection{Lateral Instability with Predecessor's Information Only}
When dealing with lateral control in convoying applications, two important issues must be addressed: (a) lateral stability and (b) lateral string stability. While stability has been discussed in previous subsections, this subsection demonstrates the inability to attenuate lateral (cross-track) errors when relying primarily on information from the preceding vehicle. Let \(e_{lat, i}(t), {\tilde \theta}_i(t)\) be the lateral and heading error respectively for the \(i^{th}\) following ACV.

\begin{definition}[\textbf{Lateral String Stability}]
A convoy of ACVs, each moving in accordance with the governing equations and control laws provided in the previous subsections, is laterally string stable if, starting with zero initial state error for each ACV, the lateral errors resulting from the lead ACV's ELC maneuver are non-amplifying, i.e., for every \(i \ge 2\)
$$\norm{\begin{pmatrix} e_{lat, i}(t) \\ {\tilde \theta}_i(t)  \end{pmatrix}}_{\infty} \le \norm{ \begin{pmatrix} e_{lat, i-1}(t) \\ {\tilde \theta}_{i-1}(t)  \end{pmatrix}}_{\infty}.$$ 
\end{definition}

The above string stability definition ensures that none of the following vehicles will be off-track if the first following vehicle remains on track. However, the following vehicles must start with zero initial errors (i.e., zero cross-track error, zero heading error, zero cross-track rate error, and zero yaw rate error), which is challenging to achieve in experiments. 

The string stability definitions in \cite{swaroop1996string} and \cite{turri2016cooperative} require bounds on initial state errors, which are not fully measurable. Consequently, we have verified string stability primarily through simulations. In experiments, determining string stability is difficult, other than checking if the following vehicles remain on-road during a maneuver.

Let \(\mathbf{x}_i\) denote the state of the \(i^{th}\) vehicle. It suffices to show lateral instability when the lead vehicle is executing a straight-line maneuver, i.e., when \(R = \infty\). With instantaneous actuation, the governing equation in \eqref{eq:System} for the \(i^{th}\) vehicle is:
\[ \mathbf{M} \ddot{\mathbf{x}}_i + \mathbf{C} \dot{\mathbf{x}}_i + \mathbf{L} \mathbf{x}_i = \mathbf{B} C_f \delta_{f,i}. \]
Taking the Laplace transformation on both sides and writing \(\mathbf{x}_i(s)\) as the Laplace transform of \(\mathbf{x}_i(t)\), we get:

\begin{eqnarray*}
&& \hspace{-1em} \mathbf{M}_0(s) := \\
&& \hspace{-1em} \begin{pmatrix}
        m_vs^2 + \frac{C_f + C_r}{V_0}s & \frac{aC_f - bC_r}{V_0}s - (C_f + C_r) \\
        \frac{aC_f - bC_r}{V_0}s & I_zs^2 + \frac{a^2C_f + b^2C_r}{V_0}s - (aC_f - bC_r)
    \end{pmatrix}, \\
&& \hspace{-1em} \mathbf{M}_0(s) \mathbf{x}_i(s) = \begin{pmatrix} 1 \\ a \end{pmatrix} C_f \delta_{f,i}(s).
\end{eqnarray*}

\(\Delta_0(s)\) is the determinant of \(\mathbf{M}_0(s)\). Since \(\det(\mathbf{M}_0(s)) = 0\) and \(\frac{d}{ds}(\det(\mathbf{M}_0(s)))|_{s=0} = 0\), it is clear that \(\Delta_0(s)\) has two roots at the origin. For some polynomial \(R(s)\), we can express \(\Delta_0(s)\) as:
\[ \Delta_0(s) = s^2 R(s). \]

With information from the preceding vehicle only, consider the control law:
\[ \delta_{f,i} (s) = -\begin{pmatrix} k_e & k_{\theta} + k_w s \end{pmatrix} (\mathbf{x}_i(s) - \mathbf{x}_{i-1}(s)). \]

Let \(\mathbf{K}(s) = C_f \begin{pmatrix} k_e & k_{\theta} + k_w s \end{pmatrix}\). Note that the components of the vector \((\mathbf{x}_i - \mathbf{x}_{i-1})\) are the lateral and heading errors with respect to the preceding vehicle. 
Note that $e_{lat, i}$ is the first component of ${\bf x}_i$ and represents the error from the straight line ($R= \infty$) the convoy is expected to track; similarly, ${\tilde \theta}_i$ is the heading error relative to the straight line the $i^{th}$ ACV is expected to track. However, the measurements available are relative errors; hence, the structure of the controller. The structure of \(\mathbf{K}(s)\) reflects the non-availability of lateral error rate information as lateral velocity information is noisy and may not be used for feedback. 

With this control law, the error propagation equation is given by:
\[ ({\mathbf M}_0(s) + {\mathbf B}{\mathbf K}(s)) {\mathbf x}_i(s) = {\mathbf B}{\mathbf K}(s) {\mathbf x}_{i-1}(s). \]

A necessary condition for string stability condition is given by:
$$\sup_{Re(s) \ge 0} \sigma_{max} (({\mathbf M}_0(s) + {\mathbf B}{\mathbf K}(s))^{-1}{\mathbf B}{\mathbf K}(s)) \le 1.$$

The above condition is equivalent to the following SISO transfer function having an infinity norm of at most one:
\[ \| \underbrace{{\mathbf K}(s) ({\mathbf M}_0(s) + {\mathbf B}{\mathbf K}(s))^{-1} {\mathbf B}}_{{\mathbf G}(s)} \|_{\infty} \le 1. \]

Let 
$$ H(s) = {\mathbf K(s)} {\mathbf M}_0^{-1}(s) {\mathbf B}.$$

Using the identity for the sum of an invertible matrix and a rank-one matrix,
$$ ({\mathbf M}_0(s) + {\mathbf B}{\mathbf K}(s))^{-1} = {\mathbf M}_0^{-1}(s) - \frac{{\mathbf M}_0^{-1}{\mathbf B} {\mathbf K}(s) {\mathbf M}_0^{-1}}{1+ H(s)},$$
we obtain
$${\mathbf G}(s)= H(s) - \frac{H(s)^2}{1+H(s)} = \frac{H(s)}{1+H(s)}.$$

The string stability condition therefore reduces to 
$$\| \frac{{\mathbf K(s)} {\mathbf M}_0^{-1}(s) {\mathbf B}}{1+{\mathbf K(s)} {\mathbf M}_0^{-1}(s) {\mathbf B}}\|_{\infty} \le 1. $$

%

For some numerator polynomial, $N(s)$, we will have
$$H(s) = {\mathbf K}(s) {\mathbf M}_0^{-1}(s) {\mathbf B} = \frac{N(s)}{s^2R(s)},$$ and consequently, 
$${\mathbf G}(s) = \frac{H(s)}{1+H(s)} =  \frac{N(s)}{s^2R(s) + N(s)} = \frac{\frac{N(s)}{R(s)}}{s^2 + \frac{N(s)}{R(s)}}.$$

We will focus on determining \(N(s)\); in particular, \(N(0)\). 
Consider the following:
$$\Delta_0(s) H(s) = \begin{pmatrix} k_e & k_{\theta} + k_w s \end{pmatrix} \Delta_0(s) \mathbf{M}_0^{-1}(s) \begin{pmatrix} 1 \\ a \end{pmatrix}.$$

From here, we can infer:
\begin{eqnarray*}
\lim_{s \rightarrow 0} s^2 H(s) \hspace{-0.75em} &=& \hspace{-0.75em} \lim_{s \rightarrow 0} \frac{\Delta_0(s) H(s)}{R(s)} \\
&=& \hspace{-0.75em}  \frac{1}{R(0)} \begin{pmatrix} k_e & k_{\theta} \end{pmatrix} \begin{pmatrix}
  -(aC_f - bC_r) & (C_f + C_r) \\
  0 & 0
\end{pmatrix} \begin{pmatrix} 1 \\ a \end{pmatrix} \\
&=& \hspace{-0.75em} \frac{(a + b) C_r}{R(0)} k_e > 0.
\end{eqnarray*}
Hence, \(\frac{N(0)}{R(0)} > 0\). Let \(\frac{N(jw)}{R(jw)} = \alpha(w) + j \beta(w)\). Then for sufficiently small \(w\), \(\alpha(w) > 0\) by continuity. 

For \({\mathbf G}(jw)\),
\[ {\mathbf G}(jw) = \frac{\alpha + j \beta}{\alpha - w^2 + j \beta}. \]
For sufficiently small \(w\) where \(\alpha > 0\), it follows that \(\alpha^2 > (\alpha - w^2)^2\). Hence,
\[ \|{\mathbf G}(jw)\| > 1 \]
at low frequencies, indicating lateral instability.

\section{Numerical \& Experimental  Results}
\label{sec:results}
\subsection{Construction of Stabilizing Set of Feedback Gains}

\begin{figure}[ht]
\centering
\includegraphics[width=0.4\textwidth]{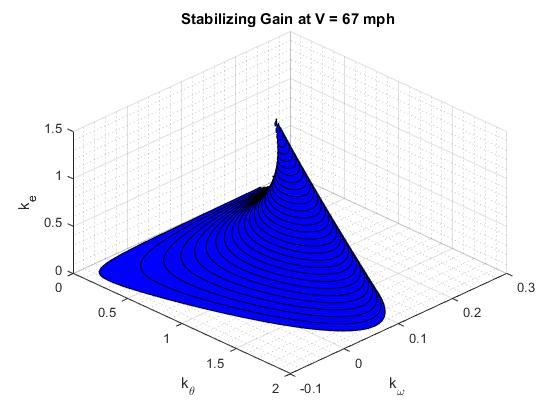}
\caption{Stabilizing set of controllers.}
\label{fig:stabilizingset}
\vspace{-0.5em}
\end{figure}

To construct the set of stabilizing feedback gains, we used the parameters from Table \ref{tab:table1}. Since the stabilizing set (which is non-convex) depends on \(V_0\), we identified the sets of all stabilizing controllers for a range of speeds \(\{10, 20, 30, 40, 50, 60, 67\} \, \text{mph}\) and selected a controller \((k_e, k_{\theta}, k_{\omega}) = (0.06, 0.96, 0.08)\) that belongs to the interior of the intersection of all these stabilizing sets. The stabilizing set corresponding to \(V_0 = 30 \, \text{m/s} \, (67 \, \text{mph})\) is shown in Fig. \ref{fig:stabilizingset}. Since the stabilizing set is bounded and the chosen set of gains is in the interior, a ball centered at the chosen controller can be entirely contained within the stabilizing set. This implies that the first condition of Proposition \ref{prop:1} is satisfied, as the eigenvalues of \(A\) are continuous functions of the controller gains, and the ball of stabilizing gains containing the chosen gains is bounded. The performance of the lateral controller with parametric uncertainty in vehicle mass will be discussed in the next subsection.

\vspace{-0.5em}
\subsection{Stability with Parametric Uncertainty in Vehicle Mass}
In the previous subsection, the set of stabilizing control gains was constructed based on the nominal mass of the vehicle. However, vehicle mass can vary depending on the number of passengers and their weight, as well as the baggage they carry. A natural question arises: Will the nominal controller maintain vehicle stability during an ELC maneuver when there are more passengers and luggage? The answer depends on whether the characteristic polynomial, given in equation (\ref{eq:cl_poly}), remains Hurwitz for the expected variation in vehicle mass.

\begin{figure}[ht]
\centering
\includegraphics[width=0.35\textwidth]{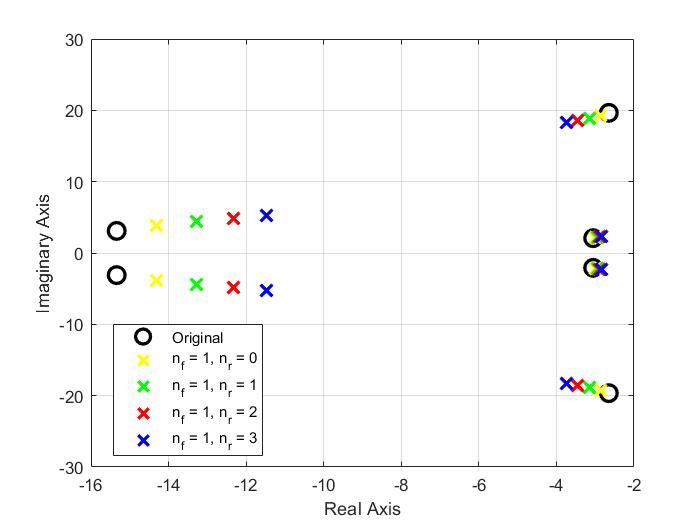}
\caption{Distribution of roots of $\Delta_o(s)$ with different vehicle masses and mass distribution.}
\label{RootsDistribution}
\vspace{-0.5em}
\end{figure}
                
Variation in vehicle mass introduces complexity in determining the location of the roots of \(\Delta_o(s)\). The moment of inertia, \(I_z\), depends not only on the vehicle mass but also on the distribution of the mass. For example, passengers may sit in the rear seats, front seats, or the baggage may be in the trunk, cabin, or on the roof. 
 
Since vehicle inertia is directly related to mass distribution, we calculate the moment of inertia as follows:
\begin{eqnarray*}
I_{update} &=& I_{nominal} + (m_{p}n_f)a^2 +\\
&&(m_{p}n_r)b^2+m_l(n_f+n_r)(b+c)^2,
\end{eqnarray*}
where \(m_p\) and \(m_l\) denote the mass of passengers and luggage, respectively; \(n_f\) and \(n_r\) denote the number of passengers in the front and rear, respectively; and \(c\) is the distance from the baggage to the rear axle. Note that the driver's mass is included in the nominal vehicle mass. Since vehicle mass directly affects the moment of inertia, the coefficients of the characteristic polynomial \(\Delta_o(s)\) are not necessarily affine in the vehicle mass parameter; hence, standard root locus arguments do not apply. However, one can numerically calculate the roots for a given vehicle mass and distribution.

For these calculations, we set \(c = 0.5 \, \text{m}\) and use the nominal gain set from the previous construction: \((k_e, k_{\theta}, k_{\omega}) = (0.06, 0.96, 0.08)\). The vehicle speed is set at \(30 \, \text{m/s} \, (67 \, \text{mph})\). We further set \(n_r \le 3\), \(n_f \le 1\), \(m_p = 70 \, \text{kg}\), and \(m_l = 50 \, \text{kg}\) for evaluating the roots of \(\Delta_o(s)\) in the characteristic polynomial shown in equation (\ref{eq:cl_poly}).

As shown in Fig. \ref{RootsDistribution}, the roots of \(\Delta_o(s)\) are located in the left half-plane in all cases, indicating the stability of the ACV's motion under parametric uncertainty in mass with the nominal lateral controller. 

\vspace{-0.5em}
\subsection{Simulation Results}

\begin{figure}[ht]
\centering
\includegraphics[width=0.35\textwidth]{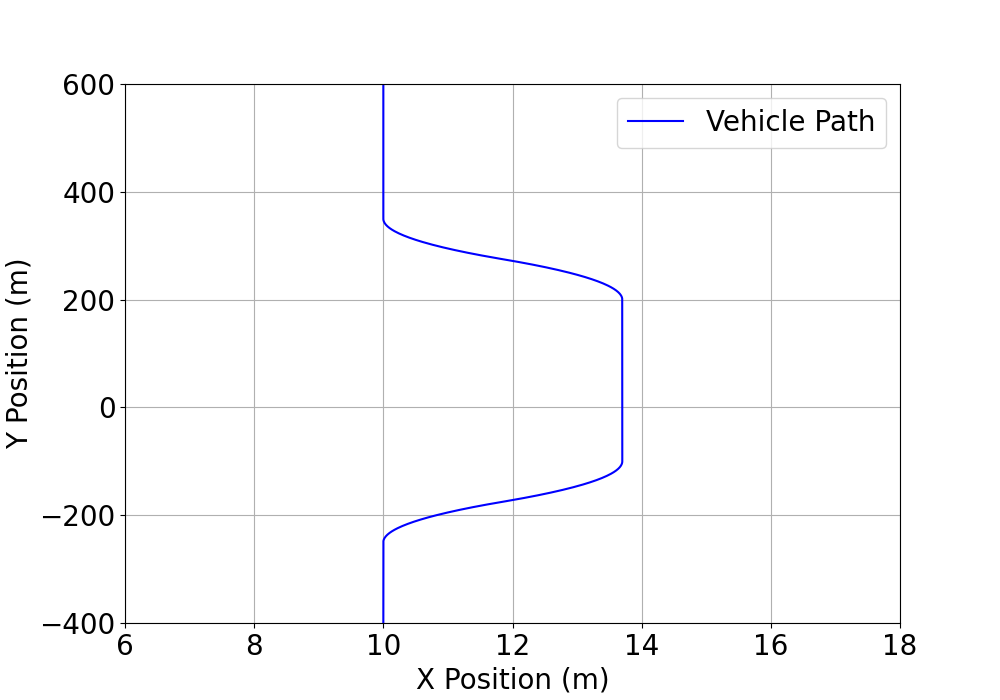}
\caption{Target trajectory for the lead vehicle.}
\label{fig:Path}
\end{figure}

Fig. \ref{fig:Path} illustrates a nominal double lane change trajectory on a 1 km road section, serving as the target trajectory for the lead ACV. This trajectory involves the ACV making a right turn to enter the adjacent right lane and then making a left turn to return to the original lane. The second ACV, as the first following ACV in the convoy, utilizes sampled trajectory information from the lead ACV and implements the control law developed in this paper. The remaining vehicles in the platoon have access to sampled trajectory information from both the lead ACV and their respective preceding ACVs, adopting the lateral controller designed in the previous section. We examine the design based on the composite ELC trajectory in different scenarios for both the lead and preceding vehicles.

\subsubsection{Lateral control performance}
To illustrate the performance of the proposed lateral control strategy, we employed a four-vehicle platoon, with each vehicle maintaining a consistent speed of \(30 \, \text{m/s} \, (67 \, \text{mph})\) and chose controller gains \((k_e, k_{\theta}, k_{\omega})\) as \((0.06, 0.96, 0.08)\). All vehicles initiated the maneuver with zero errors in their respective states as initial conditions. To assess the efficacy of the proposed lateral controller for an ELC maneuver shown in Fig. \ref{fig:Path} within a four-vehicle platoon, we considered a specific scenario:

- \(n_f = 1\), \(n_r = 3\) (one passenger in the front seat and three passengers in the rear seats).

In this scenario, each passenger was assumed to carry baggage with a weight denoted as $m_l$ in the trunk. The convoy was assumed to be homogeneous, ensuring identical mass distribution across all vehicles. For simplicity, the parameter $\alpha$ was set to 0.5 during the computation of the composite ELC trajectory for each ACV. Subsequently, based on this trajectory, feedforward control input, feedback error signals, and feedback control were determined.

\begin{figure}[ht]
\centering
\includegraphics[scale=0.25]{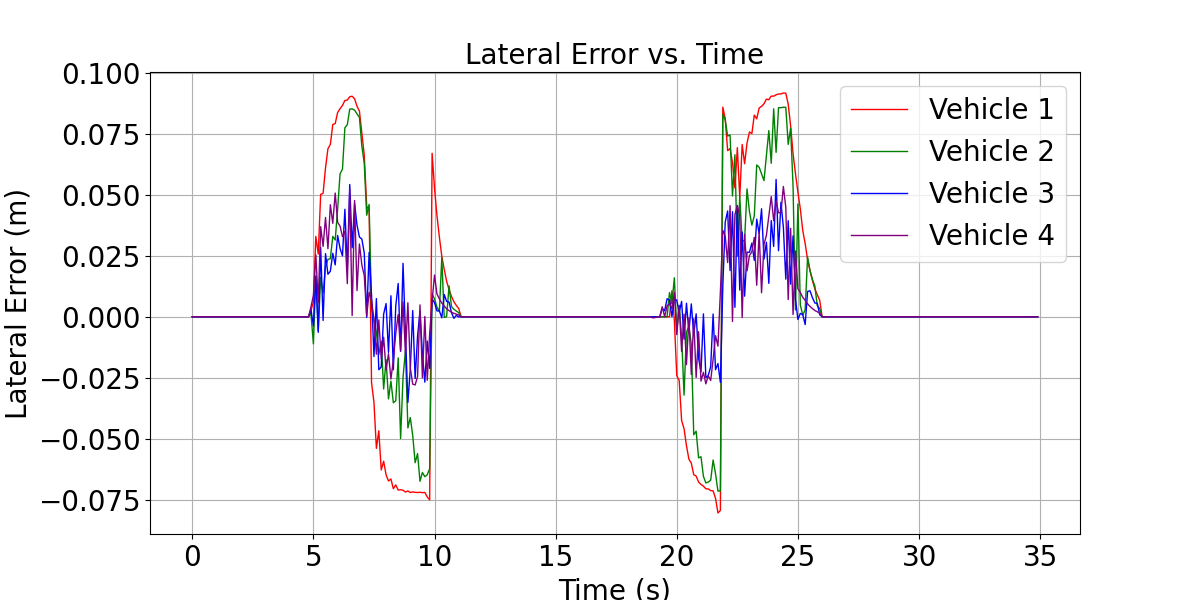}
\caption{Lateral error of vehicles with respect to composite ELC trajectory for the scenario with $n_f = 1$  and  $n_r = 3$.}
\label{Fig:LateralErrorCombineAllHeavy}
\vspace{-0.5em}
\end{figure}

Fig. \ref{Fig:LateralErrorCombineAllHeavy} illustrates the response of the control scheme in terms of lateral error while tracking the composite ELC trajectory. Notably, the double lane change maneuver commences around 5 seconds and concludes before 10 seconds, with the second lane change occurring approximately from 20 seconds to 25 seconds. The maximum lateral error less than $9 \; cm$, is observed at the initiation and termination of the curved trajectory section, as depicted in Fig. \ref{Fig:LateralErrorCombineAllHeavy}. Furthermore, the monotonically decreasing trend of the maximum lateral errors affirms lateral string stability.

\subsubsection{Contrast tests under different vehicle speeds and mass distributions}

In the preceding subsection, an analysis of the lateral controller's performance was conducted for a four-vehicle platoon, each maintaining a constant speed of \(30 \, \text{m/s} \, (67 \, \text{mph})\). The evaluation scenario involved one passenger in the front seat and three in the rear seats (\(n_f = 1, n_r = 3\)), with each passenger assumed to contribute baggage to the total mass distribution in the trunk (\(m_l\)).

Here, we aim to present simulation results for diverse scenarios encompassing different speeds and mass distributions. This serves to further illuminate the robustness and versatility of the proposed lateral controller.

Four distinct scenarios have been considered:

\begin{itemize}
   \item[(A)] Homogeneous platoon configuration: All vehicles travel at a speed of \(v = 10 \, \text{m/s}\), with each vehicle featuring one front passenger and two rear passengers (\(n_f= 1, n_r = 2\)).
   \item[(B)] Homogeneous platoon configuration: All vehicles travel at a speed of \(v = 10 \, \text{m/s}\), with each vehicle equipped solely with one front passenger (\(n_f= 1, n_r = 0\)).
   \item[(C)] Homogeneous platoon configuration: All vehicles travel at a speed of \(v = 20 \, \text{m/s}\), with each vehicle equipped solely with one front passenger (\(n_f= 1, n_r = 0\)).
   \item[(D)] Heterogeneous platoon configuration: The first two vehicles travel at a speed of \(v = 30 \, \text{m/s}\), each equipped with one front passenger and two rear passengers (\(n_f= 1, n_r = 2\)). The subsequent two vehicles, traveling at the same speed, are equipped only with a front passenger (\(n_f= 1, n_r = 0\)).
\end{itemize}

\begin{figure*}[ht!]
    \centering
    \subfigure[Scenario: A]{
    \includegraphics[scale=0.25]{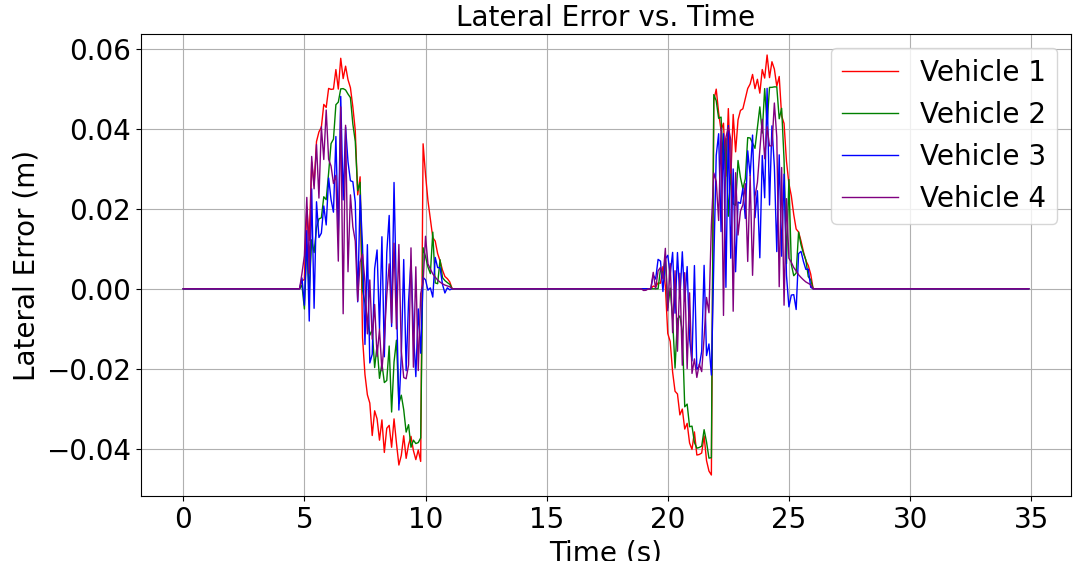}}
    \subfigure[Scenario: B]{
    \includegraphics[scale=0.252]{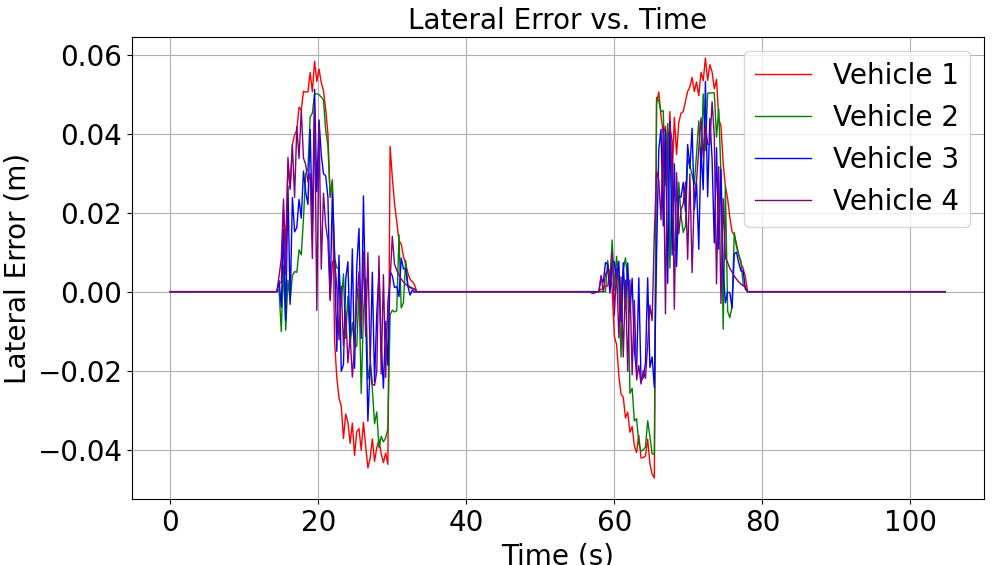}}
    \subfigure[Scenario: C]{
    \includegraphics[scale=0.252]{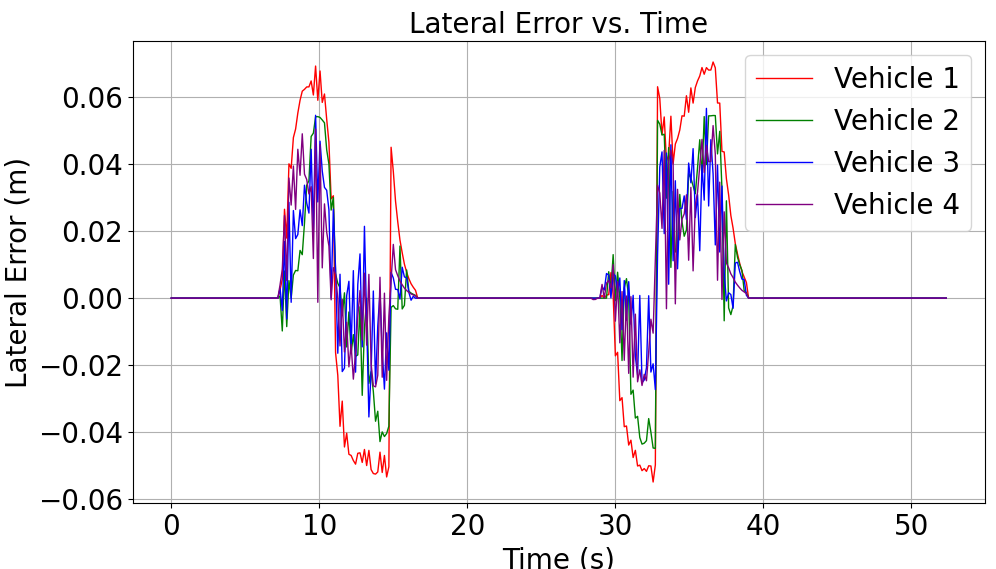}}
    \subfigure[Scenario: D]{
    \includegraphics[scale=0.25]{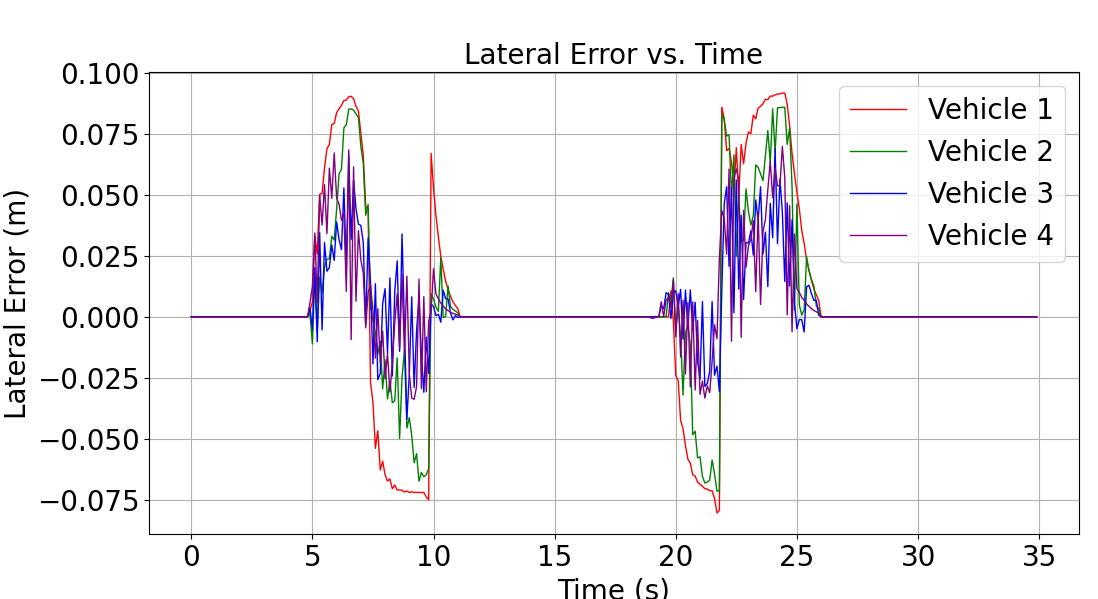}}
    \caption{Lateral error of vehicles with respect to composite ELC trajectory for different scenarios.}
    \label{Fig:Lateral_error_multiple}
\end{figure*}

In each of the four scenarios depicted in Fig. \ref{Fig:Lateral_error_multiple}, a consistent pattern of monotonically decreasing maximum lateral errors was observed. This trend validates the lateral string stability of the proposed controller across diverse operational conditions. Comparing scenarios A and C, it becomes evident that an increase in longitudinal speed leads to higher lateral errors. Nevertheless, with the correct set of control gains, lateral errors remain minimal compared to the lane width. Furthermore, comparing the first two scenarios, the negligible effect of mass distribution on the lateral errors of vehicles is apparent. Additionally, the heterogeneity of the platoon does not impact the lateral string stability of the convoy.

\subsubsection{Lateral control performance under variable speed}
\begin{figure}[H]
\centering
\includegraphics[scale=0.22]{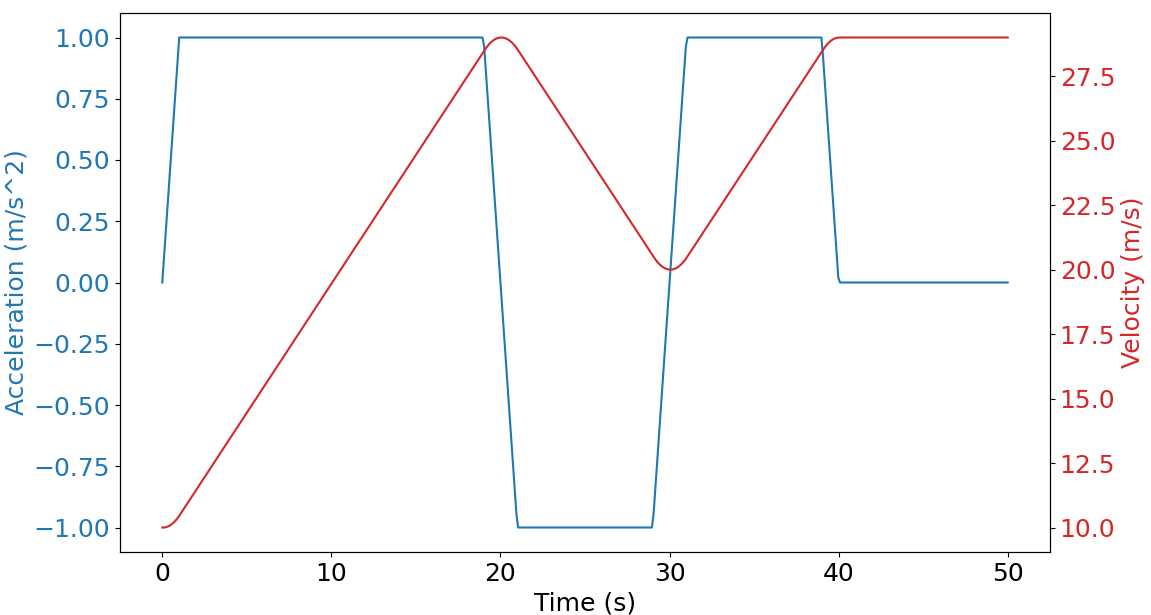}
\caption{\textcolor{black}{Acceleration and velocity profiles.}}
\label{Fig:Profile}
\end{figure}

To further validate the effectiveness of our lateral control strategy under more dynamic conditions, we extended our analysis to include variable longitudinal speed scenario. This section outlines the results of this scenario, focusing specifically on a homogeneous platoon setup where all vehicles adjust their speed according to a predetermined acceleration profile, as illustrated in Fig. \ref{Fig:Profile}.

In this scenario, each vehicle in the platoon was subjected to the same acceleration profile, designed to simulate real-world driving conditions where speeds are not constant. This speed variation was intended to test the adaptability and responsiveness of the lateral control system under changing longitudinal speed conditions.

\begin{figure}
\centering
\includegraphics[scale=0.25]{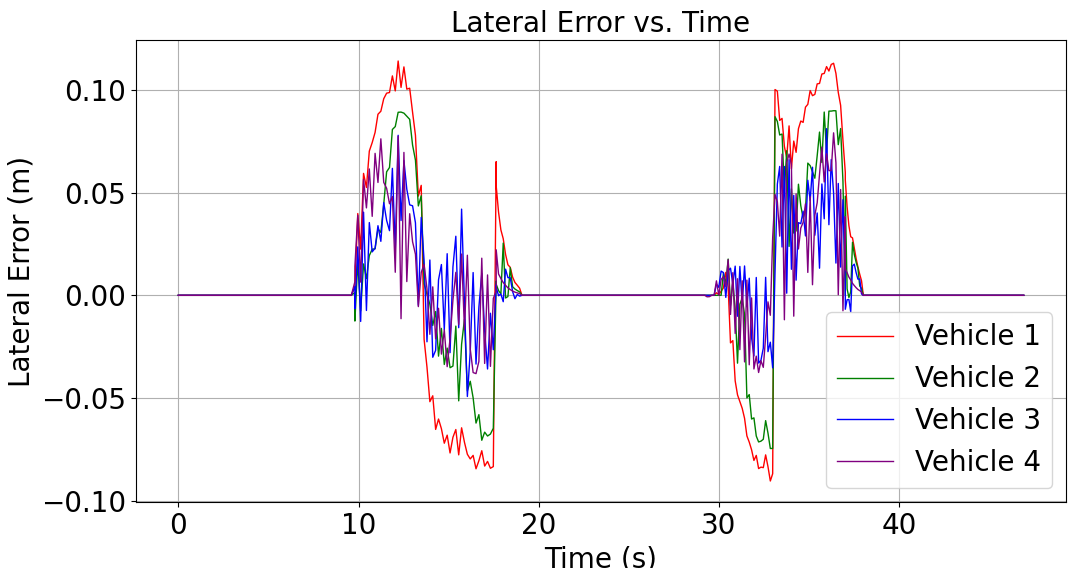}
\caption{\textcolor{black}{Lateral error of vehicles with respect to composite ELC trajectory is evaluated under a given acceleration profile.}}
\label{Fig:LateralError_Accel}
\vspace{-0.5em}
\end{figure}

Fig. \ref{Fig:LateralError_Accel} presents the lateral error measurements of the vehicles throughout the maneuver under the given acceleration profile. Notably, despite the speed variations, the maximum lateral errors remained well within the safety limits, with no errors exceeding 11 cm. This outcome demonstrates the controller's ability to effectively compensate for dynamic changes in vehicle speeds with an upper limit of \(30 \, \text{m/s}\).

The analysis showed that the lateral errors are slightly more in comparison to constant speed conditions. However, this magnitude of lateral error is minimal, especially when considered in the context of typical highway lane widths, thus affirming the practical safety of the control system. Additionally, the monotonic decrease in maximum lateral errors of vehicles in a platoon confirms the lateral string stability of the proposed control strategy.

\vspace{-0.5em}
\subsection{Experimental Results}
The control algorithms were implemented on a Lincoln MKZ car equipped with a drive-by-wire system for autonomous steering, throttle, and brake control. Vehicle state data, including position, yaw angle, and yaw rate, were obtained from an integrated GPS system and IMU unit onboard. The controller operated at a frequency of 50 Hz. To emulate a convoy, multiple runs of the same Lincoln MKZ vehicle were conducted. The control scheme allowed each following vehicle to utilize information from both its preceding ACV and the lead ACV in the convoy.

\begin{figure}
\centering
\includegraphics[scale=0.225]{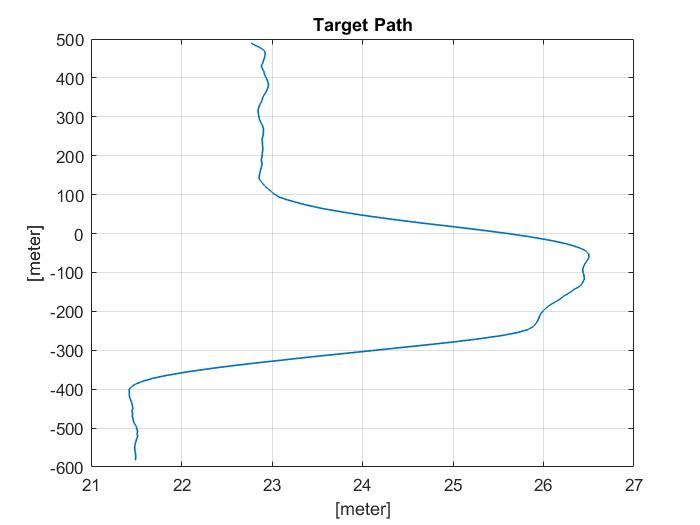}
\caption{\textcolor{black}{Vehicle reference path for experimental analysis}}
\label{fig:PathCombineExp}
\vspace{-1em}
\end{figure}

In the first run, the Lincoln MKZ vehicle served as the lead ACV, tracking the trajectory. Closed-loop trajectory information, acquired from time-stamped GPS data, was stored onboard and accessed in subsequent runs. In the second run, the vehicle accessed the stored GPS information from the first run, simulating perfectly communicated GPS data from the lead ACV. In the \(i^{th}\) run, trajectory information from both the preceding run and the first run was used to compute the steering command. During the experiment, the target speed for each vehicle in the platoon was maintained at \(25 \, \text{m/s}\). The desired trajectory is shown in Fig. \ref{fig:PathCombineExp}. Experimental analysis was conducted using the composite ELC trajectory, and the comparison of lateral errors with and without lead vehicle's data is illustrated in Fig. \ref{fig:CombinedLateralErrors}.
 
\subsubsection{String instability with lack of lead ACV's preview data}
Constructing the target trajectory for a following ACV solely based on the position data of its preceding vehicle can lead to lateral string instability. In Fig. \ref{fig:Without_lead}, \(PL1\) represents the lead vehicle, while \(PL2\) and \(PL3\) represent the first and second following ACVs, respectively. The amplification of cross-track errors in this scenario is depicted in Fig. \ref{fig:Without_lead}.

\subsubsection{String stability with lead ACV's preview data}
\begin{figure}
\centering
\subfigure[Without lead vehicle's data]{
    \includegraphics[width=0.625\columnwidth]{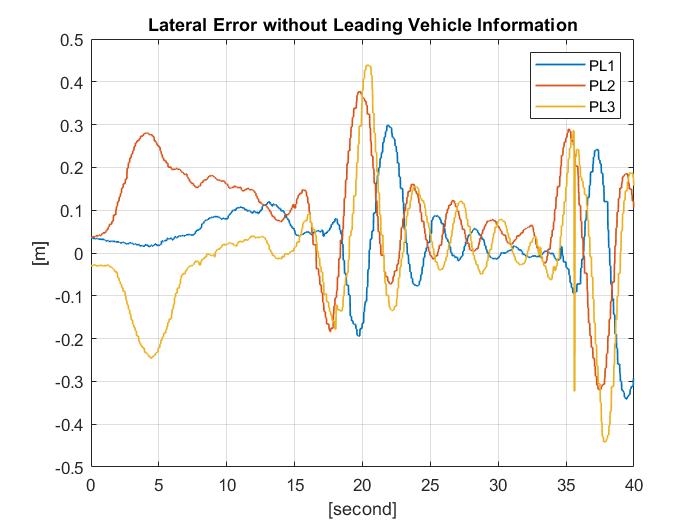}
    \label{fig:Without_lead}
}
\subfigure[With lead vehicle's data]{
    \includegraphics[width=0.615\columnwidth]{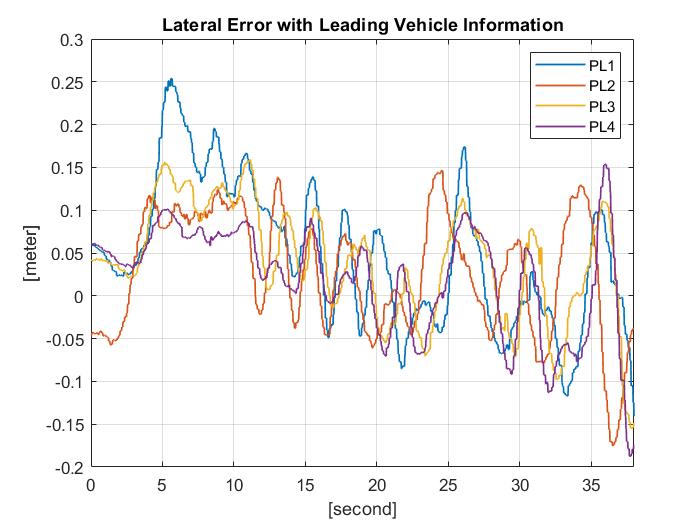}
    \label{fig:with_lead}
}
\caption{Comparison of lateral errors with and without lead vehicle's data}
\label{fig:CombinedLateralErrors}
\vspace{-1.5em}
\end{figure}

Corroborating string stability experimentally is challenging unless all the following ACVs begin with zero initial errors, as mentioned earlier. Fig. \ref{fig:with_lead} illustrates the performance of the lateral controller. The cross-track lateral errors of the following ACVs do not amplify in Fig. \ref{fig:with_lead}, in contrast to the previous case. Although the errors appear to increase at a few instances, this is primarily due to the non-zero initial error in the second vehicle, which is out of phase with the others. Nevertheless, all following ACVs successfully maintained their position on the road throughout the ELC maneuver.

\vspace{-0.75em}
\section{Conclusions}
\label{sec:conclusions}
In this work, we addressed the problem of controlling the lateral motion of a convoy of autonomous vehicles. Based on the observation of lateral string instability when using predecessor-only information, we developed a lateral control strategy for a following autonomous vehicle that utilizes sampled GPS data from both its lead and preceding vehicles. We proposed a controller architecture that effectively accommodates and fuses the communicated data to ensure stable and reliable lateral motion control. We considered data that corresponded to the position of the lead/preceding vehicle within a specified distance of the ego vehicle in the convoy. In the architecture, a composite ELC trajectory was constructed as a circular arc spline by treating both data streams as if they originate from the same source. This is simpler in terms of implementation as there is only one trajectory to be computed and only one set of feedback errors and corresponding feedback and feedforward actions to be computed. We have demonstrated that a frozen-parameter controller design can ensure closed-loop stability under certain practical conditions. These include maintaining the autonomous connected vehicle's (ACV's) longitudinal speed within a specified range—above a nonzero minimum and below a predetermined maximum—and ensuring the longitudinal acceleration is square integrable over the maneuver's duration. These stipulations are realistic, given the transient nature of ELC maneuvers. In addition to theoretical analysis, we validated the proposed control strategy through numerical simulations and real-world testing on a Lincoln MKZ vehicle. This comprehensive study confirms the effectiveness of our lateral controller in maintaining string stability across a convoy of autonomous vehicles.

\vspace{-0.75em}
\section*{DISCLAIMER}
The contents of this report reflect the views of the authors, who are responsible for the facts and the accuracy of the information presented herein. This document is disseminated in the interest of information exchange and is based on the doctoral dissertation work conducted by the first author at Texas A\&M University. The report is funded, partially or entirely, by a grant from the U.S. Department of Transportation’s University Transportation Centers Program. However, the U.S. Government assumes no liability for the contents or use thereof. 

\vspace{-0.75em}
\section*{ACKNOWLEDGMENT}
Support for this research was provided in part by a grant from the U.S. Department of Transportation, University Transportation Centers Program to the Safety through Disruption University Transportation Center (451453-19C36).


\bibliographystyle{IEEEtran}
\bibliography{references}

\end{document}